\documentclass[preprint2]{aastex}
\usepackage{natbib}

\def\var{\tablenotemark{a}}

\def\bul{$\bullet$}

\def\ddo#1{\setbox2\hbox{*}\hbox to\wd2{\hfil #1\hfil}}
\def\pr#1{\setbox2\hbox{(N)}\hbox to\wd2{\hfil #1\hfil}}
\def\no{\ddo{}\pr{ }~}
\def\nN{\ddo{}\pr{N}~}
\def\nY{\ddo{}\pr{Y}~}
\def\nS{\ddo{}\pr{\bul}~}
\def\nn{\ddo{}\pr{(N)}~}
\def\ny{\ddo{}\pr{(Y)}~}

\def\sY{\ddo{$\ast$}\pr{Y}~}

\def\sS{\ddo{$\ast$}\pr{\bul}~}
\def\so{\ddo{$\ast$}\pr{ }~}
\def\ss{\ddo{$\ast$}\pr{(\bul)}~}

\let\yY=\nY\let\yS=\nS\let\yo=\no\let\yy=\ny

\def\bt{\tablenotemark{b}}
\def\bw{\tablenotemark{b,c}}
\def\tw{\tablenotemark{c}}
\def\nc{\tablenotemark{d}}
\let\nd\nodata
\def\nddag{ ~$\cdots$\rlap{$^\dag$}~ }
\defcitealias{duqm91}{DM91}
\defcitealias{tok97}{MSC}
\defcitealias{prir05}{Paper I}

\makeatletter
\renewcommand\sidehead[1]{%
 \noalign{\vskip 1.5ex}%
 \multicolumn{7}{@{\hskip\z@}l}{#1}%
 \@ptabularcr
 \noalign{\vskip .5ex}%
}%
\makeatother
\begin{document}

\title{Contact Binaries with Additional Components  
II. A Spectroscopic Search for Faint Tertiaries}

\author{Caroline D'Angelo, Marten H.\ van Kerkwijk, and Slavek M.\ Rucinski}
\affil{Department of Astronomy and Astrophysics, University of Toronto,
    60 Saint George Street, Toronto, ON M5S 3H8, Canada}
\email{dangelo,mhvk,rucinski@astro.utoronto.ca}

\begin{abstract}
It is unclear how very close binary stars form, given that during the
pre-main-sequence phase the component stars would have been inside
each other.  One hypothesis is that they formed further apart, but
were brought in closer after formation by gravitational interaction
with a third member of the system.  If so, all close binaries should
be members of triple (or higher-order) systems.  As a 
test of this prediction, we present a search for the signature of
third components in archival spectra of close binaries.  In our sample
of 75 objects, 23 show evidence for the presence of a third component,
down to a detection limit of tertiary flux contributions of about 0.8\%
at 5200\,\AA\ (considering only contact and semi-detached binaries, we
find 20 out of 66).  In a homogeneous subset of 58 contact binaries,
we are fairly confident that the 15 tertiaries we have detected are all
tertiaries present with mass ratios $0.28\lesssim
M_3/M_{12}\lesssim0.75$ and implied outer periods
$P\lesssim10^6{\rm\,d}$.  We find that if the frequency of tertiaries
was the same as that of binary companions to solar-type stars, we
would have expected to detect about 12 tertiaries.  In contrast, if
all contact binaries were in triple systems, one would expect about
20.  Thus, our results are not conclusive, but sufficiently suggestive
to warrant further studies.
\end{abstract}
\keywords{stellar dynamics
      --- methods: data analysis
      --- binaries: close
      --- stars: formation
}

\section{INTRODUCTION}

Most stars are in binaries, yet our understanding of how these form is
far from complete (for a recent review, see \citealt{toh02}).
Particularly puzzling is the existence of very close binaries, with
orbital separations of just a few stellar radii.  That these cannot
form independently is easily seen: during the pre-main-sequence phase,
these stars would have been inside each other. Yet they exist.

  One hypothesis is that they were originally single
entities which spun up to break-up velocity during contraction and
split in two.  It is unclear, however, whether this world work: descending 
the Hayashi track, stars are centrally
concentrated and fission into roughly equal parts appears implausible.

If close binaries cannot form directly, could the stars perhaps form
as wider binaries, and be brought closer together later?  One such
possibility is that the binary is part of a hierarchical triple, and
shrink due to interaction with the third component
\citep{kis98,eggk01}.  This can work as follows: the tertiary induces
Kozai cycles \citep{koz62} in the inner binary, in which angular
momentum is transferred between the inner and outer system, leading to
cycles in eccentricity and relative inclination.  For point masses
this process is cyclical, but for stars, if the eccentricity becomes
sufficiently high, tidal effects take over at periastron, and the
orbit will circularize with a final separation of about twice the
periastron distance, i.e., much smaller than the initial one.

What helps the above is that Kozai process is weak: it only works in
the absence of anything else.  Thus, no cycles occur while there is
tidal interaction between the stars and/or their disks, i.e., as long
as the stars are young and big, nor once they have been brought in
close together.  The only requirements are that a sufficient number of
binaries be members of triple systems, and that many have sufficiently
small initial inner separations and sufficiently large relative
inclinations between the inner and outer binary planes.  None of these
constraints appear problematic: 15--25\% of all stellar systems have
three or more components \citep{tok04}, and some well-known triples
have high relative inclination (e.g., Algol; \citealt{les+93}).

Indeed the mechanism has been invoked to explain the properties of a
number of individual systems, such as the triple TY~CrA
\citep{beu+97} and the quadruple 41~Dra \citep{tok+03}.  Furthermore,
it was found by \citet{toks02} that many visual multiples had close
spectroscopic subsystems, which might have formed by the above
process.

The hypothesis has not, however, been taken to its logical conclusion:
do {\em all} close binaries form this way?  The beauty of this perhaps
far-fetched suggestion is that it makes a very clear prediction: all
close binaries should be in hierarchical triples (or higher order
systems).  In this paper, we investigate this possibility with a
particular subset of close binaries, the W~UMa contact binaries.

W UMa contact binaries -- in which the two companions share an outer
envelope -- are the closest known binaries.  While these have had yet
a further phase of orbital shrinkage, likely related to magnetic
braking and/or gravitational radiation, this phase could only happen if
they were very close binaries to start with
(\citealt{vil82,ste95}, and references therein).  Intriguingly, many
contact binaries appear to be accompanied by tertiaries: for instance,
\citet{ruck82} noted the frequent presence of visual companions, while
\citet{henm98} found the spectral signature of a tertiary in a number
of contact binaries.  Furthermore, in radial-velocity studies of
contact and other close binaries aimed at measuring their orbital
parameters, one of us (S.M.R.) has found that about one in four
binaries showed the signature of a tertiary component in its spectrum
(see \citealt{rvddo-vii} and other papers in the same series).

The above led \citet[ hereafter \citetalias{prir05}]{prir05} to
collect the available evidence for multiplicity for contact binaries
with $V<10$.  For the better-observed Northern-sky subsample, they
inferred a multiple frequency of $59\pm8\%$.  Since no method can
detect all multiples, this is a lower limit to the true fraction, but
it is difficult if not impossible to extrapolate given the complex
selection effects for various techniques and companions types.

Here we present a detailed analysis for one particular technique of
searching for the spectral signature of tertiaries.  We re-analyze the
data sets used for the radial-velocity studies referred to above
using a new technique, outlined in Sect.~\ref{sec:technique},
optimized for the detection of tertiaries.  We discuss possible
pitfalls and systematic effects, and find that these limit us
somewhat, but that we can still detect tertiaries down to flux ratios
of about~1\%, an improvement by a factor of three compared to the earlier
results.  In Sect.~\ref{sec:results}, we infer properties of
tertiaries and check consistency with previous work.  We discuss
limits and biases in our sample in Sect.~\ref{sec:analysis}, and use
the companion distribution of solar-type stars measured by
\citet{duqm91} to test the hypothesis that all close binaries are in
triple systems.  In Sect.~\ref{sec:analysis}, we also discuss what the
alternative, null hypothesis should be, i.e., what one would expect if
multiplicity had little or no influence on the formation of close
binaries; conservatively, we assume that for this case the tertiary
frequency be similar to the companion frequency of regular stars.  We
summarize our results and discuss future prospects in
Sect.~\ref{sec:conclusions}.

\section{DATA SET AND ANALYSIS TECHNIQUE}\label{sec:technique}

The data set we have available is that used in papers I-IX of a series on
``Radial Velocity Studies of Close Binaries'' (for an overview, see
\citealt{rvddo-vii}).  Briefly, it consists of almost 4000 spectra of
75 close binaries, all taken at the David Dunlap Observatory.  We list
the objects in Table~\ref{tab:sample}.  For our purposes, the
important characteristics of the spectra are that they were taken with
the $1800{\rm\,line\,mm^{-1}}$ grating centred at 5180\,\AA\ (covering
about 200\,\AA\ around the Mg I triplet), and through a 1\farcs5 or
1\farcs8 slit (matched to the typical seeing of 1\farcs7).  The
resulting slit images project to 0.64 and 0.80\,\AA, respectively,
and, including the effect of seeing, lead to an effective resolutions
of 35 to $50{\rm\,km\,s^{-1}}$.

We search for the spectroscopic signature of a tertiary by fitting the
spectrum of the contact binary, and checking whether adding a spectrum
of a fainter tertiary improves the fit significantly.  For the binary
spectrum, we can make use of the convenient fact that the contact
binary has a single spectral type (due to energy transfer from the
more massive to the less massive component, by a mechanism not
entirely understood; for a recent review, \citealt{web03}), and that
its lines are strongly broadened by the rapid rotation and orbital
motion.  In contrast, the lines of the third star should be narrower
since there is nothing to have prevented it, like all low-mass stars,
 from slowing down. (Note that there are exceptions:
 new DDO observations have revealed a broad-lined A-type companion
for V752 Mon, which entirely masks the radial velocity
 signatures of the binary so that only the variability is detectable. Since we remove 
more massive, early-type stars, this will not bias our sample.) Furthermore, 
any orbital motion of the tertiary should be small compared to that of the contact 
binary, which implies that we can analyze spectra averaged over the orbital phase of the
contact binary.  This not only increases the signal-to-noise ratio,
but should also smooth out possible relatively sharp-lined features
from the contact binary, such as might be caused by star spots.

Below we describe how we implemented the technique and how we model
the line broadening in the contact binary.  We then discuss the
criteria we use to determine whether a detection of a tertiary is
significant and real, and determine our sensitivity limits.  We
conclude with a discussion of the limitations of our method.

\begin{figure*}
\centerline{\includegraphics[width=\hsize]{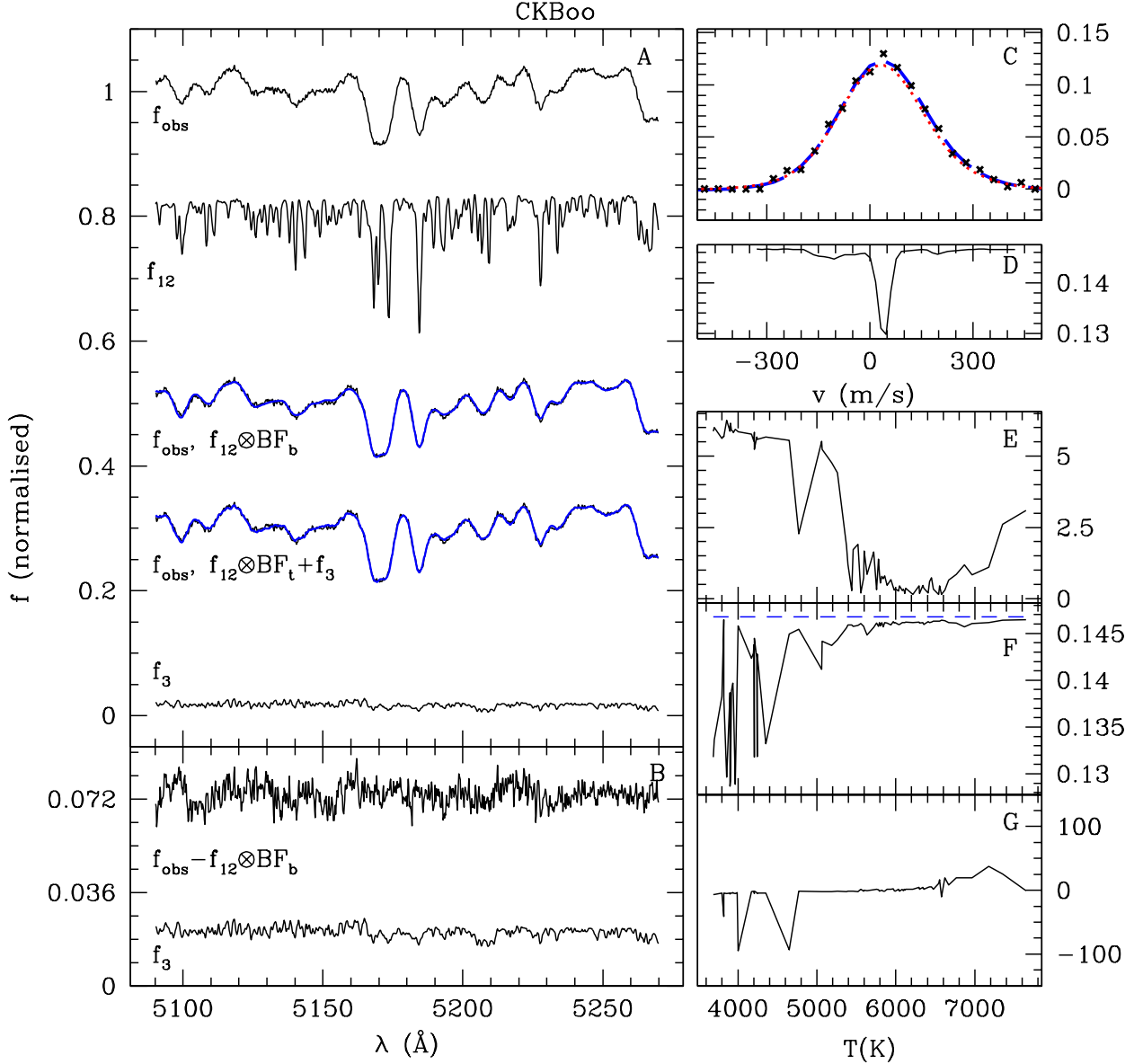}}
\caption [] {Search for a tertiary component in the spectra of the W UMa
system CK~Boo.  {\em Panel A, from top to bottom:} $f_{\rm obs}$, the
observed, average spectrum; $f_{12}$ a template of similar spectral
type used to represent the contact binary; $f_{\rm obs},~f_{12}\otimes
BF_{\rm b}$: the observed spectrum (thin line) overdrawn with the
template convolved with a best-fit broadening function (thick line);
$f_{\rm obs},~f_{12}\otimes BF_{\rm t}+f_3$: the observed spectrum
with overdrawn the best-fit model composed of the template convolved
with a re-fit broadening function, plus a tertiary spectrum; $f_3$:
the best-fit tertiary contribution. {\em Panel B:} Comparison between
the residuals from fitting the observed spectrum without including a
third star (offset by a constant value), and the best-fit tertiary
spectrum.  {\em Panel C:} The broadening function used to represent
the line profiles of the contact binary.  Crosses indicate the
empirical broadening function found from least-squares decomposition
(see Rucinski 2002), and the dashed line the fit to those points with
our three-Gaussian model shape.  This fit is used as an initial guess
for line profile; the dotted line represents the final shape, after
convergence of our procedure. {\em Panel D:} variance of the fit
residuals as a function of the velocity of the tertiary.  For CK Boo,
this shows only one minimum, which is close to the systemic velocity
of the contact binary, as expected for a real tertiary.  {\em Panels E
and F:} variance of the residuals as a function of temperature of the
template used for the contact binary and the tertiary, respectively.
For CK Boo, the best fit tertiary is substantially cooler than the
binary, as expected if it is much fainter.  {\em Panel G:} the best
fit relative velocity between the tertiary and contact binary, as a
function of tertiary temperature.  For CK Boo, this is close to zero,
as expected for a physically associated component.\label{fig:outline}}

\end{figure*}


\begin{figure*}
\centerline{\includegraphics[width=\hsize]{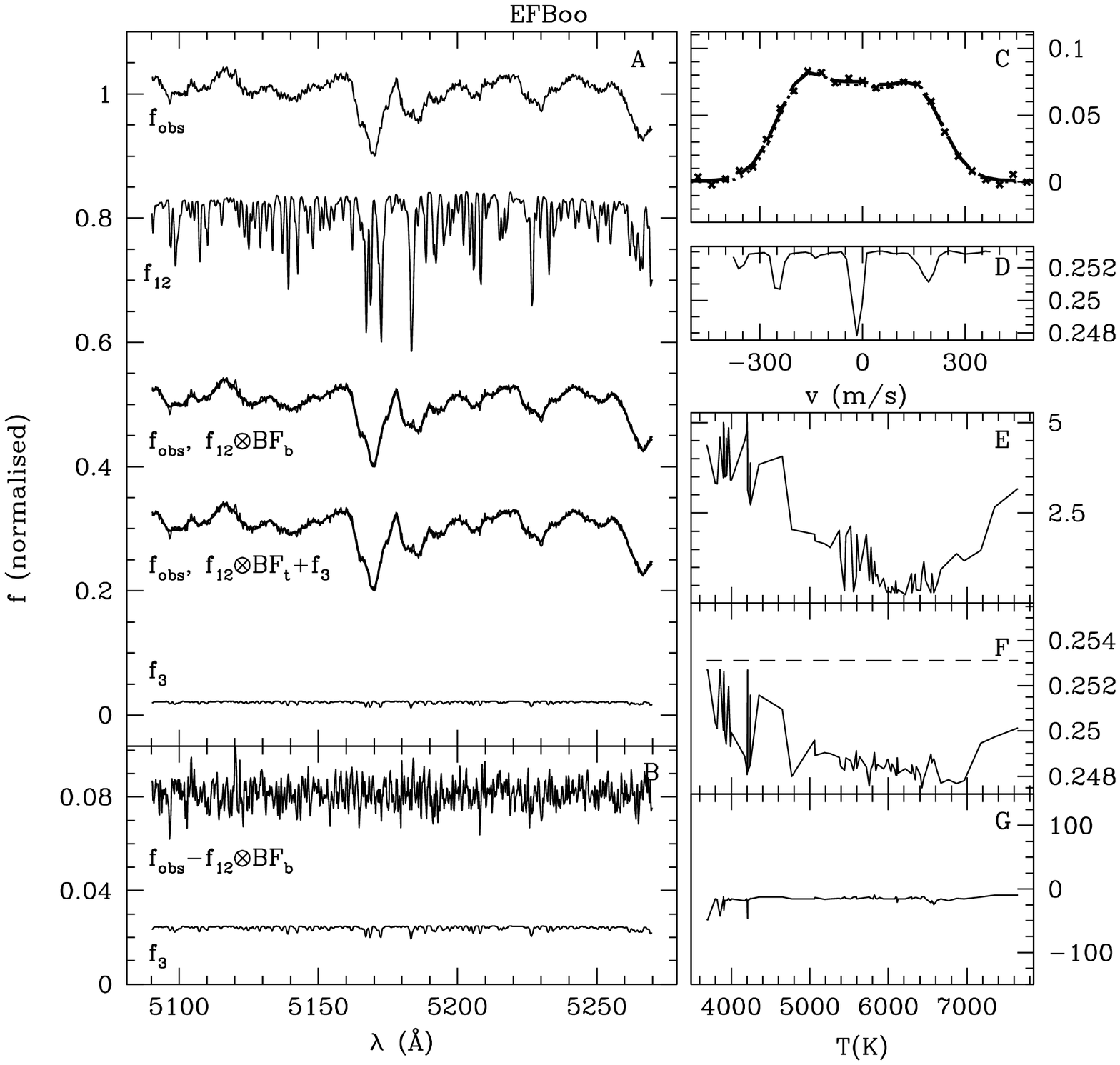}}
\caption{Like figure~\ref{fig:outline}, but for the W UMa system
EF Boo ($f_{obs}$), in which we do not detect a third star down to a level of 0.8 $\%$.}
\end{figure*}

\subsection{Implementation}

To search for tertiaries, we fit the data using the procedure outlined
in Fig.~\ref{fig:outline}.  Starting with the average of all
normalized, barycentered spectra of a given source, we try to
reproduce it with a template spectrum of a sharp-lined, slowly
rotating star convolved with a model broadening function optimized to
match the binary's line profile.  For our templates, we use the
database of high-resolution, normalized stellar spectra obtained with
ELODIE (\citealt{prus01}), repeating our fit procedure for all spectra
in order to find the one that matches best.  The model broadening
function is composed of three Gaussians with equal width.  We choose
such a model to ensure that our broadening function is wide enough that the
sharp-lined signal from a putative tertiary is not removed; since this
was a particularly difficult part of our analysis, we describe it in more
detail in Sect.~\ref{sec:bf} below.


To measure the quality of the fit, we convolve the model spectrum with
a truncated Gaussian (to simulate the effects of seeing and transfer
through the slit), regrid on the observed pixel array, multiply with a
polynomial function to simulate differences in the normalization, and
finally calculate the variance between the observations and model.  We
minimize the variance as a function of the various parameters using
the Downhill Simplex method as described by \citet[ \S10.4]{pre+92}.

Once the best fit to the contact binary spectrum has been determined,
we add a third star to the spectrum, optimize the relative flux and
velocity, and determine the resulting improvement in the fit.  We
again repeat this procedure for a wide range of different spectral
types (from early F down to early~M).

With our procedure, the final set of parameters determined is: (i)
best-fit template spectrum for the contact binary, or, more
interestingly, its temperature; (ii) the systemic velocity; (iii) the
width, separation, and two relative intensities for the three-Gaussian
broadening profile (see Sect.~\ref{sec:bf}); (iv) the best-fit
template for the third star; (v) its fractional intensity; and (vi)
its velocity relative to the contact binary.  In addition, there are
up to 10 parameters without physical meaning, viz., those that
describe the polynomial accounting for difference in continuum
normalization.

\subsection{The Broadening Profile}\label{sec:bf}

We found that the most difficult part of our analysis was to accurately reproduce the line-profile shape of the contact binary.  In
principle, with good phase coverage, one might expect that a single
Gaussian would suffice.  In practice, however, the binaries were
observed preferentially near the quadratures.  The result, for systems
with mass ratio near unity, is a double-humped line profile in the
average spectrum, while for systems with extreme mass ratios the
profile shows a prominent central hump (from the more massive
component), as well as two `side lobes' (from the less massive one).

Based on this structure, we chose to model the broadening profile with
a set of three Gaussians, but constrained to have identical width and
separation.  This leaves four free parameters: the width, the
separation, and the relative intensities of the two outer Gaussians
(since the whole profile is normalized, the intensity of the central
Gaussian does not have to be specified).

Another problem that arose in automating our procedure was that the
profiles of the different systems were so varied that it was
impossible to use a single set of initial guesses for the broadening
profile that worked for all systems.  To circumvent this, we first
determined empirical broadening profiles at the instrumental
resolution by least-squares decomposition (using the technique
described in detail by \citet{rvddo-vii}; see Fig.~\ref{fig:outline},
panel C).  Next, we fit these empirical broadening profiles
using our three-Gaussian model, and used the resulting parameters as
initial guesses for our main procedure.

\begin{figure*}
\centerline{\includegraphics[width=0.5\hsize]{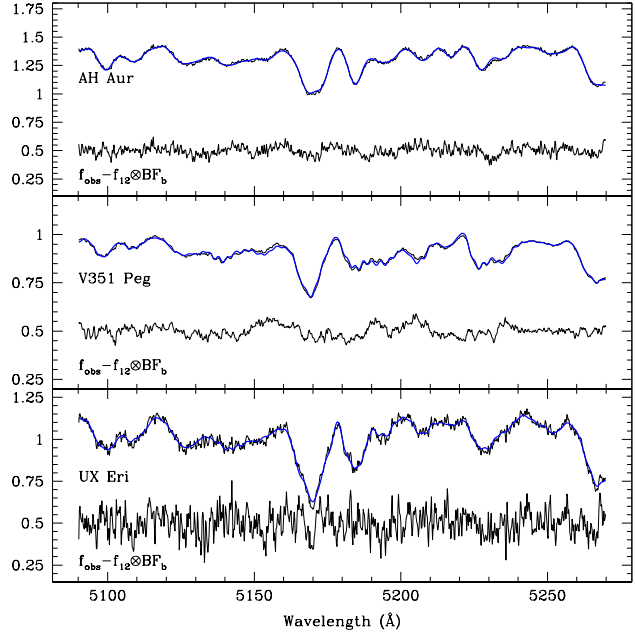}}
\caption{Three examples of problems encountered in fitting the contact binary, which
make detecting a third star more difficult. Top: AH Aur and its
best fit (in bold overlay), with the residuals (enlarged) below. A
poor match in spectral type to the contact binary leads to large-scale
residuals. Middle: V351 Peg and the residuals from its best fit. In
this case, the dominant residuals come from errors in fitting the
contact binary's complicated broadening profile, and so show up as high frequency
residuals. Bottom: UX Eri and the residuals from its best fit. In this
case, the spectrum itself is noisier (in addition to the poor spectral
match) and so the residuals are larger. In general, even with these
effects we can detect a third star down to a level of about
$\beta \simeq 0.8 \%$; for the worst cases (which we exclude in our analysis) this level is closer to
$\beta \simeq 2 -3 \%$.\label{fig:systematics}}
\end{figure*}

\subsection{Detections and Detection Limits}\label{sec:detections}

With the fit results in hand, we need to determine whether or not a
possible improvement resulting from adding a tertiary is significant.
For some objects, this is trivial: the residuals from the fit with the
binary model show a clear signature of a different spectral type.
Those, however, would typically have been found already in the earlier
studies, since the tertiary component would lead to a narrow peak in
the empirical broadening function (see \citealt{rvddo-vii}).

For fainter tertiaries, one could in principle use statistical tests
to determine whether the improvement in variance (or, equivalently,
$\chi^2$) is significant.  This only works, however, if the variance
is dominated by measurement noise.  In practice this is not the case:
the quality of the fit is usually limited by a mismatch between our
model and the true contact binary's spectrum.  Indeed, even for our
spectra with the worst signal-to-noise (such as UX Eri;
Fig.~\ref{fig:systematics}, bottom panel), we find that systematics
dominate.  As a result, the quality of the fits is not good in a
statistical sense, and the use of $\chi^2$ becomes meaningless.

In Fig.~\ref{fig:systematics}, we show the types of more severe
systematic mismatches that limit our sensitivity to tertiaries.  The
first is a poor match to the binary's spectral type, which happens
mostly for cooler temperatures for which the ELODIE archive contains
relatively few suitable templates.  As can be seen for the case of AH
Aur in Fig.~\ref{fig:systematics} (top panel),
the mismatch leads to low-frequency residuals and thus an increased
variance.  Since the residuals have long wavelength, one could still
tease out the signature of a tertiary, but since the limits will not
be as good, we decided not to include objects like AH Aur in our
statistical sample.

The second source for systematic error is more problematic: poor
matches to the binary star's line profile.  For cases such as V351~Peg
(Fig.~\ref{fig:systematics}, middle panel), where our three-Gaussian
broadening function does not match the intrinsic profile very well,
high-frequency residuals are left.  In consequence, there is an
obvious danger of a false detection of a ``third star'' that matches
these residuals.

In order to avoid the above pitfalls, we decide whether or not a
detection is significant using not only visual inspection of the
residuals, but also the following two physical arguments.  First, for
a faint tertiary, the temperature should be substantially lower than
that of the contact binary, and, therefore, as a function of tertiary
temperature, minimum variance should occur at low values.  Second, the
orbital motion of the tertiary should be relatively small, and hence
as function of relative velocity, minimum variance should occur near
zero.  In the example shown in Fig.~\ref{fig:outline}, both criteria
are met, and hence we consider the detection of the tertiary secure.
There are also, however, a fair number of sources for which tertiary
flux and the decrease in variance are similar, but for which the
inferred tertiary temperature is higher than that of the contact
binary and/or the radial velocity is inconsistent.

With the above procedure, we find that we are able to detect
tertiaries down to fluxes of 0.8\% of that of the contact
binary.  We confirmed this by adding third stars at different flux
levels to objects for which we did not detect tertiaries, and finding
the level at which we could recover those: we found a limit of 0.9\% even for
cases where the match to the line profile was relatively poor, such as
UX Eri and V351 Peg.

\subsection{Limitations}

Apart from the problems addressed above in obtaining an adequate fit
to the binary, our technique also has limitations inherent in the
assumptions we make about the tertiary.  An obvious one is that we
cannot detect compact objects, since these would be too faint and
likely not contribute any spectral features.  Another is that we
have few templates for late-type stars.  This is not an issue for the
more massive, earlier contact binaries, for which such late-type
tertiaries would be undetectable.  But for later-type contact
binaries, we might miss cool tertiaries or, more likely, overestimate
their temperatures and fluxes (the latter since the strength of the
band heads, etc., generally increases with decreasing temperature).

A different limitation arises from our assumption that the tertiary
rotates slowly.  For early-type tertiaries, this may not be correct.
Those, however, would be very bright and hence noted independently
(furthermore, we will exclude them from our sample since we cannot be
sure the sample of contact binaries with such bright tertiaries is
complete; see Sect.~\ref{sec:biases}).  For late-type tertiaries, slow
rotation is expected unless the star is in a close binary and is kept
corotating by tidal forces.  Thus, our procedure will not identify
close binaries as companions (such as the quadruple system composed of
two contact binaries, BV Dra and BW Dra; \citealt{ruck82}).  In order
for the projected rotation velocity to be below our resolution, i.e.,
$v\sin i\lesssim50{\rm\,km\,s^{-1}}$, one requires $P_{\rm
rot}\gtrsim1{\rm\,d}$ (for a $1\,R_\odot$ star).  But at such short
orbital periods, orbital velocities would be even higher, and those
would smear the signal as well (at least for systems for which the
spectra were obtained over an extended period of time).  Orbital
velocities for the tertiary were indeed found in HT Vir
\citep{rvddo-iv}.  In order not to decrease our sensitivity, a
tertiary that is itself a binary should have a radial-velocity amplitude
$K\lesssim50{\rm\,km\,s^{-1}}$, which requires $P_{\rm
orb}\gtrsim20\,$d (for two $1\,M_\odot$ stars).


\begin{deluxetable}{llll@{~~~~~~}llll@{~~~~~~}llll}
\tablewidth{0pt} 
\tabletypesize{\footnotesize}
\tablecaption{Sample of Close Binaries\label{tab:sample}}
\tablehead{%
\colhead{Star}&\colhead{Ref.}&\colhead{$\beta$\var}&\colhead{Notes\bt}&
\colhead{Star}&\colhead{Ref.}&\colhead{$\beta$\var}&\colhead{Notes\bt}&
\colhead{Star}&\colhead{Ref.}&\colhead{$\beta$\var}&\colhead{Notes\bt}}
\startdata
AB And&     IX&   \nd& \nY   & DK Cyg&     II&   \nd& \nn BF& V753 Mon&  III&   \nd& \nN   \\
CN And&    III&   \nd& \nN NC& V401 Cyg&   VI& 0.015& \yy   & V502 Oph&   IX&   \nddag& \nS   \\
GZ And&      I& 0.015& \sY TW & V2082 Cyg&  IX& 0.020& \sS   & V839 Oph&   II&   \nd& \nY   \\
V376 And&    V&   \nd& \nN   & V2150 Cyg&  IV&\nddag& \nY   & V2357 Oph&VIII&   \nd& \no   \\
EL Aqr&      V&   \nd& \no   & RZ Dra&    III&   \nd& \no NC& V2377 Oph&  IV&   \nd& \nN   \\
HV Aqr&    III& 0.022& \sS   & BX Dra&     IX&   \nd& \no BF& V2388 Oph&  VI& 0.103&  \yY   \\
V417 Aql&    I&   \nd& \no   & GM Dra&     VI&   \nd& \nN   & V1363 Ori&  IX&   \nd& \nn   \\
AH Aur&     II&   \nd& \ny BF& FU Dra&    III&   \nd& \no   & BB Peg&      I& 0.009& \ss \\
V402 Aur& VIII&   \nd& \nN   & SV Equ&     II&   \nd& \no NC& KP Peg&     IX& 0.03&  \yY NC\\
V410 Aur& VIII& 0.22&  \yY   & UX Eri&    III&   \nd& \ny   & V335 Peg&   IX&   \nd& \nN   \\
44 Boo\tw& IV& 0.23&   \yY TB& QW Gem&   VIII&   \nd& \ny   & V351 Peg&    V&   \nd& \nN BF\\
CK Boo&     II& 0.009& \sY   & V842 Her&   II&   \nd& \nN   & AQ Psc&      I&   \nd& \nY   \\
EF Boo&      V&   \nd& \nN   & V899 Her&   IV& 0.725&   \yS TB& DV Psc&     IV&   \nd& \no NC\\
FI Boo&     IV& 0.012& \sY   & V918 Her&   IX&   \nd& \nN   & OU Ser&    III&   \nd& \nN   \\
SV Cam&     VI& 0.016& \so NC& V921 Her& VIII&   \nd& \nN   & EQ Tau&      V&   \nd& \nn   \\
AO Cam&    III& 0.008& \sS   & V972 Her&   VI&   \nd& \nN   & V1130 Tau&VIII&   \nd& \no NC\\
DN Cam&      V&   \nd& \nY   & FG Hya&      I&   \nd& \nN   & HN UMa&   VIII&   \nd& \nN   \\
FN Cam&      V&   \nd& \nN   & UZ Leo&     II&   \nd& \nN   & HX UMa&   VIII& 0.023& \yY   \\
V523 Cas& VIII&   \nd& \ny BF& XZ Leo&     II&   \nd& \ny   & II UMa&     VI& 0.148&  \yY   \\
V776 Cas&    V& 0.015&  \yY TW& ET Leo&     VI& 0.022& \sS   & GR Vir&     II&   \nd& \nN   \\
V445 Cep&   IX& 0.055& \sY   & EX Leo&     IV&   \nd& \nN   & HT Vir&     IV& 0.282&  \yY   \\
EE Cet&     VI&\nddag& \nY   & FS Leo&     VI&   \nd& \no NC& KZ Vir&      V&   \nd& \no   \\
KR Com&     VI& 0.23&  \yY TB& RT LMi&    III&   \nd& \no   & NN Vir&     II&   \nd& \nN   \\
YY CrB&    III&   \nd& \nN   & VZ Lib&     IV& 0.045& \ss   & HD 93917\tw&VIII&\nd& \nN  \\
SX Crv&      V&   \nd& \nN   & SW Lyn&     IV& 0.194&  \yo NC& NSV 223\tw& VIII&\nd& \no  
\enddata
%
%
\tablenotetext{a} {  ${\rm \beta}$ is the flux ratio ($\equiv f_3/f_{12}$)
determined from our fit (${\rm{}\beta}\la\!0.008$ for non-detections;
further properties for the detected systems are listed in
Table~\ref{tab:properties}. A brief description can be found in the Appendix
for all systems with detections as well as some interesting triple systems we missed, marked with $\dag$.)}

\tablenotetext{b} {Notes  
  {\bf{*}}:the spectroscopic signature of the tertiary was not recognized in the original DDO series paper; 
  {\bf{Y}}, {\bf{\bul}}, {\bf{N}}: the system is in the sample of \citetalias{prir05}. parentheses indicate the extended, $V_{\rm max}>10$ sample: {\bf{Y}}--identified as a triple independently from the DDO spectra,{\bf{\bul}}--identified
  as a triple based on the DDO spectra,{\bf{N}}--not identified as a triple;
  {\bf{NC}}, {\bf{BF}}, {\bf{TB}}: the system was not included in our statistical analysis
  because it is not in contact (i.e., the temperatures of the two
  components differ), because our procedure yielded a bad fit to its
  spectrum and hence our sensitivity to tertiaries was poor, or because
  the tertiary is too bright to be sure the sample is complete; {\bf{TW}}:
  the system is included in our statistical analysis, but the detection
  of the tertiary is not counted because the tertiary is at
  too wide a separation, and we cannot be sure we could detect such
  tertiaries for all objects in our sample.}
\tablenotetext{c}{Variable names: 
  44 Boo=i Boo; 
  HD 93917=VY Sex; 
  NSV 223=DZ Psc}
\tablerefs{I: \citet{rvddo-i}; 
II: \citet{rvddo-ii};
III: \citet{rvddo-iii};
IV: \citet{rvddo-iv};
V: \citet{rvddo-v};
VI: \citet{rvddo-vi};
VIII: \citet{rvddo-viii};
IX: \citet{rvddo-ix}.}
\end{deluxetable}

\section{RESULTS}\label{sec:results}

Out of 75 systems, we have detected tertiaries for 23, nine of which had
been missed in the original analysis of the spectra.  All detections
are indicated in Table~\ref{tab:sample} and described in more detail in
the Appendix.  Below, we will compare our results to those in the
literature, and then proceed to infer tertiary masses and mass ratios.
A summary of observed and inferred properties of the triple systems is
given in Table~\ref{tab:properties}.

\subsection{Comparison to Previous Results}\label{sec:previous}

In Table~\ref{tab:sample}, we indicate for all stars in our sample
whether it was also studied in \citetalias{prir05}, and, if so,
whether there was independent evidence for it being member of a triple
(or higher-order) system.  For many of our detections
there is independent evidence for multiplicity, and, conversely,
for most systems for which we detect no tertiary, there is little
evidence to the contrary.  This likely reflects the fact that
most methods, whether detecting companions through gravity or flux,
require relatively similar minimum masses.

There are nine systems, however, for which a tertiary was found in
\citetalias{prir05} but not in our analysis.  For five cases -- EE
Cet, V2150 Cyg, QW Gem, AQ Psc, and AH Aur -- the discrepancy is
simply that the separation is too large for any light of the tertiary
to have entered the slit (for EE Cet and V2150 Cyg, some light did
enter the slit, which, knowing that they were visual binaries, we
could detect; see the Appendix).  For three others -- AB And, V523
Cas, and UX Eri -- the minimum mass inferred from the arrival-time
variations implies a flux below our threshold (which is the case for
$M_3/M_{\rm CB}\lesssim0.28$ [Sect.~\ref{sec:inferred}]; V523 Cas also
was fit particularly poorly [Sect.~\ref{sec:biases}]).  For the
remaining system -- DN Cam -- the identification is based on suspected
multiplicity from Hipparcos and X-ray emission in excess of
expectations.  Since this does not yield a mass estimate, we cannot
check whether our non-detection makes sense.

Turning now to the properties of the systems in which we detect
tertiaries (Table~\ref{tab:properties}), we see that the
contact-binary temperatures inferred from our spectral fits (column
$T_{\rm fit}$) are generally in fair agreement with those inferred
from \bv\ ($T_{B-V}$, found by interpolation in Table~15.7 of
\citealt{cox00}).  This gives confidence in our method.  There are three
exceptions: KR Com, 44 Boo and V899 Her, in both of which the tertiary
contributes a significant fraction to the system's light.  As a result, an 
incorrect temperature for the contact binary is found in the first step 
of our procedure.  Since this temperature is not checked {\em a posteriori}, 
it is expected that the other results will also be inaccurate (we do not 
use these systems in our statistical analysis; see Sect.~\ref{sec:biases}).

Comparing the flux ratios ${\rm \beta}_{\rm fit}\equiv f_3/f_{12}$ from
our fit to those from the literature (${\rm \beta}_{\rm lit.}$), there are also 
a number of discrepancies.  For resolved systems, the literature
values should be reliable, and hence we need to understand what went
wrong in our procedure.  We see two possible causes for errors.  First, as above,
for systems with bright tertiaries (44 Boo and KR Com) the temperature
assigned to the contact binary will be incorrect and hence the other
parameters will be inaccurate.  Second, for systems with wide
separations (V776 Cas, KP Peg, and, to a lesser extent, 44 Boo), the
tertiary would have been only partially in the slit, and hence the
flux will be underestimated.

Turning now to unresolved triple systems, which were all identified in
the DDO program and thus have values based on the same data, we find
that our flux ratios are systematically smaller, especially for
fainter tertiaries.  In the DDO program, the tertiaries were
recognized by the appearance of a sharp feature in the broadening
function at the system's average radial velocity, and the flux ratio
was determined from the ratio of the area under the sharp peak to the
remainder of the broadening function (\citealt{rvddo-vi}).  Since
these broadening functions are derived using a least-square
decomposition based on a single template spectrum, the
contribution of the tertiary was effectively measured under the assumption that it
had the same spectral type as the contact binary.  If its true
spectrum has stronger lines -- as will be the case for faint
tertiaries with cooler temperatures -- this will lead to an
overestimate of its contribution.  Since our procedure uses a separate
spectral type to derive the contribution from the tertiary, our flux
ratios should be more reliable.

The above issues allow one to understand the discrepancies between
literature values and those derived here, but make it difficult to
estimate reliable uncertainties.  From the comparison with resolved
systems, uncertainties of $\lesssim\!15\%$ are indicated for flux
ratios between 0.05 and 0.5, but errors increase towards higher and
lower values.  For the brighter tertiaries, we will use flux ratios
from resolved observations, which should be good to $<\!10$\%.  For
fainter ones, however, one needs to keep in mind that our
uncertainties increase rapidly, reaching $\sim\!50\%$ for ratios below
$\sim\!0.02$.

\begin{deluxetable}{lllllllllllllllll}
\tablewidth{0pt}
\rotate
\tabletypesize{\footnotesize}
\tablecaption{Observed and Inferred Properties for Close Binaries with
Tertiaries.\label{tab:properties}}
\tablehead{&
\multicolumn{7}{c}{\dotfill Contact Binary\dotfill}&
\multicolumn{7}{c}{\dotfill Tertiary\dotfill}\\
\colhead{Star}&
\colhead{$P$}&\colhead{$M_2/M_1$}&\colhead{\bv}&
\colhead{$T_{\rm fit}$}&\colhead{$T_{B-V}$}&
\colhead{$M_V$}&\colhead{$M_1$}&
\colhead{$\beta_{\rm lit.}$}&\colhead{sep.}&
\colhead{$\beta_{\rm fit}$}&
\colhead{$T_{\rm fit}$}&\colhead{$T_{\rm \beta}$\tablenotemark{a}}&
\colhead{$M_V$\tablenotemark{a}}&\colhead{$M_3$}&
\colhead{$M_3/M_1$}&\colhead{$M_3/M_{12}$}\\
&\colhead{(d)}&&&\colhead{(K)}&\colhead{(K)}&&\colhead{($M_\odot$)}&
&\colhead{(\arcsec)}&&\colhead{(K)}&\colhead{(K)}&&\colhead{$(M_\odot)$}&}
\startdata
GZ And&    0.305& 0.514& 0.79& 5600&    5200& 4.80& 0.96& 0.011$\ddag$& 2.13&  0.015&  3500&    4000& 9.69& 0.42& 0.43& 0.29\\
HV Aqr&    0.374& 0.145& 0.63& 5800&    5800& 3.92& 1.22& \nd&   \nd&   0.022&  4000&    4000& 8.06& 0.59& 0.48& 0.42\\
V410 Aur&  0.366& 0.144& 0.56& 5400&    6000& 3.75& 1.29& 0.38$\dag$&  1.7&   0.22&  5200&    5600& 4.80& 0.97& 0.75& 0.66\\
44 Boo&    0.268& 0.487& 0.94& 5400\bt& 4800& 5.50& 0.85& 2.08$\dag$&  1.7&   0.23\bw&  6100\bt& 5953& 4.70& 0.99& 1.17& 0.79\\
CK Boo&    0.355& 0.111& 0.54& 6600&    6100& 3.75& 1.29& 0.007$\ddag$& 0.12&  0.009&  3900&    3900& 9.14& 0.47& 0.37& 0.33\\
FI Boo&    0.390& 0.372& 0.64& 5800&    5700& 3.87& 1.24& \nd&   \nd&   0.012&  3900&    3800& 8.67& 0.52& 0.42& 0.31\\
AO Cam&    0.330& 0.415& 0.58& 5800&    5900& 4.01& 1.18& \nd&   \nd&   0.008&  4200&    4000& 9.25& 0.46& 0.39& 0.28\\
V776 Cas&  0.440& 0.130& 0.47& 6500&    6500& 3.12& 1.49& 0.238$\dag$& 5.38&  0.015\tw& 6100&    5600& 4.68& 1.00& 0.67& 0.59\\
V445 Cep&  0.449& 0.167& 0.12& 7400&    8400& 2.03& 1.95& \nd&   \nd&   0.055&  5600&    6600& 5.18& 0.91& 0.46& 0.40\\
KR Com&    0.408& 0.091& 0.52& 6100&    6200& 3.42& 1.42& 0.58$\dag$&  0.119& 0.23&  6100&    5800& 4.01& 1.19& 0.84& 0.77\\
V401 Cyg&  0.582& 0.290& 0.3 & 6700&    7300& 2.07& 1.93& 0.03&  \nd&   0.015&  4700&    4700& 6.63& 0.73& 0.38& 0.29\\
V2082 Cyg& 0.714& 0.238& 0.31& 7000&    7200& 1.71& 2.14& \nd&   \nd&   0.020&  5100&    5200& 5.95& 0.79& 0.37& 0.30\\
V899 Her&  0.421& 0.566& 0.48& 6300\bt& 6400& 3.24& 1.44& 1.5&   \nd&   0.73\bt&  6400\bt& 6500& 3.59& 1.36& 0.94& 0.60\\
ET Leo&    0.347& 0.342& 0.61& 5800&    5800& 4.00& 1.19& \nd&   \nd&   0.022&  3900&    3900& 8.15& 0.58& 0.49& 0.36\\
VZ Lib&    0.358& 0.237& 0.61& 5800&    5800& 3.94& 1.21& 0.2&   \nd&   0.045&  4700&    4200& 7.31& 0.67& 0.55& 0.45\\
V2388 Oph& 0.802& 0.186& 0.41& 6100&    6800& 1.78& 2.10& 0.19$\dag$&  0.088& 0.10&   5900&    5900& 3.59& 1.36& 0.65& 0.55\\
BB Peg&    0.362& 0.360& 0.52& 5900&    6200& 3.65& 1.33& \nd&   \nd&   0.009&  4000&    3900& 8.76& 0.51& 0.39& 0.28\\
HX UMa&    0.379& 0.291& 0.44& 6600&    6700& 3.32& 1.44& 0.047$\dag$& 0.63&  0.023&  4400&    4400& 6.64& 0.73& 0.50& 0.39\\
II UMa&    0.825& 0.172& 0.4 & 6600&    6800& 1.70& 2.15& 0.23$\dag$&  0.87&  0.15&   6100&    6400& 3.29& 1.45& 0.67& 0.58\\
HT Vir&    0.408& 0.812& 0.56& 6100&    6000& 3.54& 1.23& 0.586$\S$& 0.6&   0.28&    6100&    5900& 4.12& 1.15& 0.93& 0.52\\
SV Cam\nc& 0.593& 0.641& 0.62& 5800&    5800& 3.00& 1.49& \nd&   \nd&   0.016&  3900&    3900& 7.49& 0.65& 0.44& 0.27\\
SW Lyn\nc& 0.644& 0.524& 0.38& 7200&    6900& 2.12& 1.87& 0.33&  \nd&   0.19&   6200&    6500& 3.90& 1.23& 0.66& 0.43\\
KP Peg\nc& 0.727& 0.322& 0.06& 7400&    8900& 0.92& 2.68& 0.52$\dag$&  3.5&   0.03\tw& 7700&    6600& 1.63& 2.19& 0.82& 0.62\\
\enddata
\tablenotetext{a}{For resolved triples, $\beta_{\rm lit.}$ was used to
  infer $T_{\rm \beta}$ and $M_V$; for all others, $\beta_{\rm fit}$ was
  used.}  
\tablenotetext{b}{The values inferred from our fit are inaccurate,
  since the tertiary is brighter than the contact binary and dominates
  the average spectrum.} 
\tablenotetext{c}{The flux ratio is inaccurate, since only a fraction of
  the tertiary's light fell inside the slit.} 
\tablenotetext{d}{SV Cam, SW Lyn, and KP Peg are not contact
  binaries.  Therefore, the components do not have equal temperature
  and the deduced properties are less reliable.  They are not used in
  our statistical analysis.} 
\tablerefs{The binary periods ($P$), mass ratios ($M_2/M_1$) and $\bv$ are taken
from the original papers I-IX (see Table~\ref{tab:sample}).  $\beta_{lit.}$ and the separation angle
are from the same sources except where indicated:
$\dag$: \citealt{tok97};
$\ddag$: \citetalias{prir05};
$\S$: \citep{hei86}.}
\end{deluxetable}

\subsection{Inferred Properties}\label{sec:inferred}

Contact binaries follow a period-luminosity-color relation, which
allows one to derive the absolute magnitude $M_V$ from the period $P$
(in days) and dereddened color $(\bv)_0$ (\citealt{ruc04}, and
references therein), 
\begin{equation}
M_V = -4.44\log P + 3.02(\bv)_0 + 0.12.
\label{eq:plc}
\end{equation}
With the tertiary flux ratio, this yields the absolute magnitude of
the tertiary.  Next, we use the fact that contact binaries are on the
main sequence, and that, therefore, the generally fainter tertiary
should be on the main sequence as well.  Then, with the main-sequence
mass-luminosity relation (we use \citealt{cox00}, Tables 15.7 and
15.8), the tertiary mass $M_3$ follows from the absolute V-band
magnitude.  The results of this procedure are listed in
Table~\ref{tab:properties}.  Here, we did not correct for the
typically very small reddening ($E_{B-V}=0.00\ldots0.03$; Rucinski,
unpublished work).  The resulting errors in the absolute magnitude
inferred from Eq.~\ref{eq:plc} are on the order of 0.1\,mag,
substantially below the 0.25\,mag scatter in the period-luminosity
relation.\footnote{Ignoring reddening leads to a small systematic
underestimate of the brightness and thus the mass of the tertiary.
Since the binary's mass will be underestimated as well, however, the
effect on the mass ratio should be small.}

The uncertainty in the derived masses has contributions from all
steps.  The magnitudes predicted from the period-luminosity-color
relation are uncertain by $\sim\!0.25\,$mag \citep{ruc04}.  For
relatively bright tertiaries, the uncertainty in the tertiary flux
ratio is smaller ($\lesssim\!15\%\,$, or 0.15 mag).  With a total
uncertainty in $M_V$ of $\sim\!0.3\,$mag, this leads to an uncertainty
in the derived masses of $\lesssim\!10$\%.  Evolution along the main
sequence will likely contribute less, except for the brightest
tertiaries.  For fainter tertiaries, the uncertainty in the flux ratio
leads to an error in $M_V$ of as much as 0.7\,mag, but since the
mass-luminosity relation becomes much steeper for fainter objects, we
still expect the final uncertainty in the mass to be around
$\sim\!10$\%.

In order to verify the above, we also derived tertiary temperatures
from $M_V$ (column $T_{\rm \beta}$).  These should be similar to those
inferred from the spectral fits ($T_{\rm fit}$); from
Table~\ref{tab:properties}, one sees that this is indeed the case.  We
note, however, that this is not a strong test, since temperature does
not vary strongly with stellar mass.\footnote{For the same reason, it
is not very useful to infer masses from the fitted temperatures.}

Finally, in order to derive mass ratios, we also need an estimate for
the masses of the stars in the contact binary.  For this purpose, we
use that the contact binary's luminosity will be the sum of the
luminosities of its two roughly main-sequence components.  Since the
luminosity depends steeply on mass, this implies that for low mass
ratios $q\equiv M_2/M_1$, the luminosity is simply that of the
primary, while for higher ones there is a contribution from the
secondary.  To estimate the primary's absolute magnitude, $M_{V,1}$,
we use that for main-sequence stars, the V-band luminosity scales
as $M^{4.4}$ (inferred from \citealt{cox00}, Table~15.7 and 15.8; for
a more detailed analysis, see \citealt{moc81}), so that
\begin{equation}
  M_{V,1} \simeq M_V+2.5\log\left(1+q^{4.4}\right)
\end{equation}
Next, we estimate the primary's mass $M_1$ using
the main-sequence mass-luminosity relation (\citealt{cox00}, Tables
15.7 and 15.8).  Including the uncertainty in how far the star has
evolved on the main sequence, we expect these masses to be accurate to
$\lesssim\!20$\%.

\section{STATISTICAL ANALYSIS}\label{sec:analysis}

In order to determine whether the number of tertiaries we find is
consistent with the hypothesis that all close binaries form in triple
systems, we need to ensure our sample is homogeneous.  Thus, we need
to remove systems for which our procedure did not work properly, and
consider for which separations and masses (or mass ratios) we can be
certain we would have detected a tertiary if one were present.  We
flag all systems we exclude from our sample in Table~\ref{tab:sample} and discuss our reasons in more detail below. Next, we compare
our results for tertiaries with those found for secondaries for
solar-type stars, trying to extrapolate the tertiary frequency to
masses and separations to which we are not sensitive, and testing the
hypothesis that all contact binaries are in triple systems.

\subsection{Limits and Biases in our Sample}\label{sec:biases}

Among the sample of close binaries observed at DDO most are contact
binaries, but nine are not: CN And (somewhat uncertain;
\citealt{rvddo-iii}), SV Cam, RZ Dra, SV Equ, FS Leo, SW Lyn, KP Peg,
DV Psc and V1130 Tau.  For these, the assumption of a single spectral
type for both stars is inappropriate and hence our procedure will not
work optimally.\footnote{We believe the detections of tertiaries in SV
Cam, SW Lyn, and KP Peg are reliable despite the fact that these
systems are not in contact.}  We thus exclude all nine systems in our
statistical analysis.

We exclude a further five systems because the fit to the binary's
spectrum was too poor to detect a third star down to flux ratios of
0.008.  For one of these, V523 Cas, the temperature is low and we do
not have a good template in our library.  For a further four, AH Aur,
DK Cyg, BX Dra, V351 Peg, the match to the line profile is poor.

We now turn to physical limits and biases.  First, since our method is
based on spectra taken through a $1\farcs8$ slit, we will only be able
to detect tertiaries at relatively close separations.  For separations
in excess of $\sim\!1\farcs8$, the contribution of tertiary light will
be reduced and hence we will only detect very bright objects (such as
V776 Cas).  At a typical distance of $\sim\!100{\rm\,pc}$, and
including a statistical correction factor of $10^{0.13}=1.35$ (as in
\citealt{duqm91}) for projection effects, this corresponds to a
separation of $\sim\!240{\rm AU}$ or, assuming a total mass of the
system of $\sim\!2\,M_\odot$, an orbital period of
$\sim\!2600{\rm\,yr}\simeq10^6{\rm\,d}$.

Second, a tertiary needs to be sufficiently bright.  For main-sequence
stars, we cannot detect tertiaries with flux ratios below
$\sim\!0.008$.  From Table~\ref{tab:properties}, one sees that this
corresponds to mass ratios $M_3/M_{12}\simeq0.28$.  The value does not
appear to depend much on the properties of the contact binary.  To see
why, we consider three possible configurations spanning the extremes
of the range of contact-binary properties seen, with masses
$M_1=\{1,1,2\}\,M_\odot$ and $M_2/M_1=\{0.3,0.8,0.3\}$.  For those
parameters, the absolute magnitudes $M_{V,12}\simeq\{4.7,4.4,2.0\}$
and a tertiary with, e.g. $\beta$=1\% would be at
$M_{V,3}\simeq\{9.7,9.4,7.0\}$.  This corresponds to tertiary masses
$M_3\simeq\{0.42,0.45,0.70\}\,M_\odot$, and thus to mass ratios
$M_3/M_{12}\simeq\{0.32,0.25,0.27\}$.  We conclude that we could
detect all tertiaries with mass ratios in excess of~0.28.

Third, independent of our method, we must somehow be able to observe 
a contact binary.  For very bright tertiaries, the contact 
binary would be completely outshone and it likely will not be 
detected unless the tertiary is a star that appears interesting on 
its own accord and is studied in detail.  But there is a bias even 
if the tertiary is less bright, contributing, say, only half the flux.  
In such a case, the variability would still be detectable, but the system 
might well be misclassified, since the narrow lines 
of the tertiary would stand out in the spectrum 
while the broad ones from the contact binary would be
much harder to detect (a good example of such a system is TU UMi;
\citealt{ruc+05}).  Given the above, we expect the sample of known
contact binaries to be biased against systems having tertiaries
brighter than the contact binary (which, roughly, corresponds to a
mass in excess of that of the primary, or a mass ratio $M_3/M_{12
}\gtrsim0.75$).

Finally, more generally, since we rely on the spectroscopic signature
of a tertiary, we cannot detect white dwarfs, neutron stars or black
holes.

In summary, out of a sample of 75 close binaries, there are 61 contact
binaries for which our method worked well, among which we detect 20
tertiaries.  Among these, however, three (44 Boo, V899 Her and KR Com) should
not be included, since the tertiary is at least half as bright as the binary and we
cannot be confident our sample of contact binaries is unbiased for
such systems.  Furthermore, V776 Cas and GZ And should not be counted as tertiaries, since the
separation between the binary and third star is too large, making us incomplete. 
Thus, we are left with a sample of 58
contact binaries, for which we can be reasonably confident that our 15
detected tertiaries constitute all main-sequence tertiaries with $0.28\lesssim
M_3/M_{\rm{}CB}\lesssim0.75$ and $P_3\lesssim10^6{\rm\,d}$.

\subsection{Comparison with Solar-type Binaries}

Our method will miss triple systems with tertiaries that are at large
separations, have low mass, and/or are compact.  To estimate
their number we need to extrapolate, but we do not know {\em a
priori} the mass-ratio and separation distributions of the tertiary.
By way of an estimate, we will assume that these are the same as those
found for binary companions to solar-type stars.  Along the way, we
will try to test the alternate hypothesis that not all contact
binaries have tertiaries, but that rather that the companion frequency
is similar to that of solar-type stars.

Before proceeding, we note that our choice of alternate hypothesis is
somewhat arbitrary.  One would like to test the hypothesis that the
number of tertiaries we find is consistent with what one expects from
the formation of multiples.  This number, however, is not known:
multiplicity among very young stars is very poorly constrained, and
even among older stars there is no complete census (for a status
report, see \citealt{tok04}).  Our choice of comparing with solar-type
stars corresponds to an implicit assumption that the companion
frequency is independent of whether or not an inner system is a single
star or a binary.  It seems likely that this is a conservative
assumption, i.e., assuming multiplicity plays no role in the formation
of close binaries, the companion frequency of contact binaries is
unlikely to be higher than that for solar-type stars.

For the solar-type stars, we use the \citet[ hereafter
\citetalias{duqm91}]{duqm91} sample of 164 stars of spectral type F7
to G9 (masses $\sim\!0.8$ to $1.3\,M_\odot$).  Using their
distributions of mass ratio and period, we will mimic the selection
effects present in our sample.  Before doing so, however, two
complications need to be mentioned.  One is that among the 81 orbits
used in the mass ratio and period distributions, six are second orbits
from triple systems,\footnote{\citetalias{duqm91} mention seven triple
systems, but do not use the extremely wide outer orbit of the triple
HD 122660.}  and four are second and third orbits from two quadruple
systems.  For our purpose of estimating probabilities of finding a
companion with certain parameters, this leads to an overestimate
(e.g., for the full \citetalias{duqm91} sample, the number of single
stars is 93, not $164-81=83$).  To avoid biasing ourselves, we treat
the \citetalias{duqm91} sample as consisting of 174 targets (some of
which are stars, some binaries, some triples), 81 of which have a
companion.

A second complication is that the sample of \citetalias{duqm91} is
divided into two groups: spectroscopic and resolved (visual and common
proper motion) binaries, which correspond to binaries with periods
$P<10^4{\rm\,d}$ and $P>10^4{\rm\,d}$, respectively.  These groups
differ in that white dwarfs will be present among the (single-lined)
spectroscopic binaries, but not among the resolved ones.  In their
statistical analysis of mass ratios, \citetalias{duqm91} inserted
eight `fake' white-dwarf systems in the long-period group: two each in
the four mass bins with $q_{\rm max}=\{0.5,0.6,0.7,0.8\}$.  We will
remove these, since we cannot detect white dwarf companions.  We will
also correct for the presence of white dwarfs among the spectroscopic
binaries.

We now turn to the application of the selection effects present in our
sample.  First, to mimic the incompleteness among known contact
binaries, we ignore all binaries with mass ratios $q\equiv
M_2/M_1>0.75$: from lines (3) and (5) in Table~7 of
\citetalias{duqm91}, after correction for one `fake' white dwarf in
line (5), we find this reduces the sample by 6.15 spectroscopic and
7.5 resolved binaries.  (Here, the numbers are non-integer since we
had to split the $q=0.7$--0.8 bin and since for the spectroscopic
binaries, \citetalias{duqm91} corrected for the distribution of
orbital inclinations).  This leaves a sample of 160.35 targets.

Second, to reproduce our detection limit, we count all binaries with
mass ratio $q>0.28$ among the remaining targets: again from lines (3)
and (5), after correction for seven `fake' white dwarfs, we find 22.27
and 31.8 binaries, respectively.  

Third, to account for our separation limit, we select systems with
periods $P<10^6{\rm\,d}$.  For this purpose, we use that from the
period distribution (Fig.~7 in \citetalias{duqm91}), among the
resolved systems with $P>10^4{\rm\,d}$, 31 out of 65 have
$P<10^6{\rm\,d}$.  Thus, only 15.17 out of the 31.8 long-period binaries
remain, and the total implied detection rate for a survey like ours
would be $(22.27+15.17)/160.35=23\%$.

In the above, we still need to correct for the presence of white
dwarfs among the spectroscopic binaries.  \citetalias{duqm91} mention
that from statistical considerations of stellar populations, one
expects ``about two white dwarfs per decade of period.''  This would
imply about eight are present among their sample of 34 spectroscopic
binaries.  Seven of these would be included in the 22.27 binaries with
$0.28<q<0.75$ selected above, implying a reduced detection rate of
19\%.  This may be an overestimate.  On the other hand, the selected
\citetalias{duqm91} sample includes a number of companions with
periods below $10{\rm\,d}$, which could not exist around contact
binaries.\footnote{One of these, 44 Boo, is in our sample as well.}
Since these numbers are no more than guesses, we will use a rounded expected detection rate
of 20\% for our analysis below.

In summary, we conclude that if companions to contact binaries
occurred at the same frequency as those to solar-type stars, and if
their properties followed the same mass-ratio and period
distributions, we should have detected tertiaries for about 20\% of
our sample, or 12 out of 58.  In reality, we found 15 tertiaries.
This is 3 more than expected, but the difference is not highly
significant: there is a 17\% probability to find 15 or more tertiaries
out of 60 systems when the expected tertiary rate is 20\%.

To calculate a similar probability for our hypothesis that all close
binaries are in triple systems, we need to estimate the probability
that a tertiary will have the correct properties to be detected with
our method.  For this purpose, we use again the \citetalias{duqm91}
sample and first estimate the companion fraction for all systems with
$q<0.75$, again by adding up lines (3) and (5) in Table~7 (we thus
include the white dwarfs).  Without the highly uncertain $q<0.1$ bin,
we find 34.75 spectroscopic and 59 resolved binaries, respectively, or an implied companion
fraction of $93.75/160.35=58$\%.  If we include the estimated 5.6
spectroscopic and 14 resolved binaries with $q<0.1$, this rises to
67\%.  Thus, between 30 and 34\% of all companions have $q>0.28$,
$P<10^6{\rm\,d}$, and are not white dwarfs.  If the tertiary rate were
100\%, these would be the expected detection rates, and hence we would
expect to have found between 17 and 20 tertiaries in our sample of 58.
Conservatively assuming the expected tertiary rate is 34\%, we find
that there is a 12\% probability of finding 15 or fewer systems.  Thus,
our results are also consistent with the hypothesis that all close
binaries are in triple systems.

Finally, for all our above estimates we assumed that the tertiaries
followed the same mass-ratio and period distributions as those of
solar-type companions.  We do not have sufficient objects to test this
rigorously, but can at least verify this hypothesis.  For the ranges
$q=\{[0.28,0.4\rangle,[0.4,0.5\rangle,[0.5,0.6\rangle,[0.6,0.75\rangle\}$,
we found $\{8,3,3,1\}$ tertiaries.  Consulting Table~7 of
\citetalias{duqm91}, and scaling to the same total number of companions (15), we infer
 $\{6.4,3.2,3.5,1.9\}$ 
binaries (where, as above, we reduced the long-period bin by a factor $31/65$ to
correct for periods $>\!10^6{\rm\,d}$ and deducted one system in the
three higher mass-ratio bins in order to correct for white dwarfs
among the spectroscopic binaries).  Clearly, within the limited
statistics, the two distributions are consistent.

\section{CONCLUSIONS AND PROSPECTS}\label{sec:conclusions}

We searched a sample of 75 close binaries for the spectroscopic
signature of tertiaries and identified 23 triple systems, implying a
ratio of almost one in three.  For a homogeneous subset of 58 contact
binaries, we are fairly confident our 15 tertiaries are all those that
have periods $\lesssim\!10^6{\rm\,d}$ and mass ratios $0.28\lesssim
M_3/M_{\rm CB}\lesssim0.75$.

We compared our results with expectations under two hypotheses, that
the incidence of tertiaries is similar to the incidence of companions
to solar-type stars, and that all close binaries are in triple
systems.  The latter hypothesis is expected to hold if close binaries
form via the Kozai mechanism; the former appears a conservative upper
limit for the case that the formation of close binaries is unrelated to multiplicity.
Using the \citetalias{duqm91} sample of companions to solar-type stars
to infer mass-ratio and period distributions, we find that, for the
two hypotheses, the expected numbers of triple systems among our
sample are 12 and 20, respectively.  Finding 15 systems is consistent
with either hypothesis.

While inconclusive in terms of testing the role of multiplicity in the
formation of close binaries, we feel the relatively large fraction of
triple systems found is encouraging, especially since in Paper I, from
a variety of methods, a high tertiary fraction of $59\pm8\%$ was observed as well.
 To make progress, a larger
fraction of tertiary parameter space needs to be covered.
For the method presented here, this is not difficult since the archive observations
used here were not optimized for the search for faint tertiaries.  By
using higher resolution spectra spanning a larger wavelength range,
one can improve the contrast in the spectra and increase the
sensitivity, and by observing at longer wavelength, one will be
sensitive to lower mass tertiaries for a given limiting contrast
ratio.  All three improvements are possible with echelle
spectrographs.  Furthermore, with adaptive optics in the near
infrared, one can reach even lower mass tertiaries (though only on
relatively long orbits).  We hope to follow both routes in the future.

\acknowledgements We thank Peter Eggleton and Andrei Tokovinin for enlightening
discussions, and Stefan Mochnacki and Fang Bao for help in the initial
stages of this project.  We made extensive use of the SIMBAD database
and the VizieR catalogue access tool, both operated at CDS,
Strasbourg, France, and of NASA's Astrophysics Data System.  We
acknowledge financial support by NSERC.

\appendix

\section{INDIVIDUAL SYSTEMS}

Below, we briefly summarize the properties of all binaries for which
we detected the spectroscopic signature of a tertiary, as well as for
a number of systems without detections for which the results warrant
further discussion (marked with $\dag$ in flux ratio column, $\beta$, of
Table~\ref{tab:sample}).  The roman numeral directly following the
system's name is the paper in the series of ``Radial-Velocity Studies
of Close Binary Stars'' from which our data were drawn; see the
references at the end of Table~\ref{tab:sample}.  For data on resolved
multiples, we generally rely on the Multiple Star Catalogue (MSC, June
2005 update; \citealt{tok97}).

\paragraph{GZ And} (I) is a W UMa binary, and the brightest
component of a visual multiple system.  Despite the noisiness of the
spectrum and the relatively poor match for our best temperature of
5600K, we clearly detect a third companion, with $\beta=0.015$
and $T_3=3500{\rm\,K}$.  Of all visual components, the only one that this
could correspond to, is component E of \citetalias{prir05}, which is
at a separation of $2\farcs13$ and has $\Delta H=2.6$, $\Delta K=2.4$.
Using Table~\ref{tab:properties} and \citet[ Tables 7.6, 15.7, and
15.8]{cox00}, we infer $M_K=3.3$ for GZ~And.  Thus, component E has
$M_K\simeq5.7$ and is likely an $\sim\!0.42\,M_\odot$ M2 star with
$T_3\simeq3500\,$K, $V-K\simeq4.0$ and $M_{V,3}\simeq9.7$.  The
temperature agrees very well with our measurement, and hence we are
confident we detected component~E.  The implied flux ratio of
$\beta\simeq0.011$ is somewhat smaller than what we measure, the
opposite of what is expected given that the tertiary would not have
been completely in the slit.  Likely, our measurement is biased by the
relatively poor fit.

\paragraph{HV Aqr} (III) is an A-type contact binary system, and is
one of the best examples of our program's ability to detect components
at flux ratios of only a few percent.  Although visually there appears
little improvement between the fit with and without a third star and
it is difficult to see the third star's contribution in the residuals
from the binary-only fit, there are distinct minima in the variance as
a function of $T_3$ and $\Delta v_{3}$.  Furthermore, the fitted
temperature, $T_{\rm3,fit}=4000\,$K, matches that inferred from the
flux ratio, $T_{\rm3,\beta}=4000\,$K.  In \citetalias{prir05}, no other
indicators for multiplicity were found.

\paragraph{V410 Aur} (VIII) is a W UMa-type binary in a known triple
system, with a tertiary at $1\farcs7$ that is fainter by $\Delta
V=1.04$ \citepalias{tok97}, corresponding to $\beta=0.38$.
The signal of the tertiary is obvious in the spectra and from its
contribution to the broadening function, a flux ratio of 0.26 was
inferred, while from our routine we infer 0.22.  Likely, both numbers
are lower than the true flux ratio because some of the light fell 
outside the slit.  The temperature $T_{\rm3,fit}=5200\,$K is 
consistent with that inferred from the flux ratio, $T_{\rm3,\beta}=5600\,$K.

\paragraph{44 Boo B} (IV) is the contact binary nearest to Earth, and
its spectrum is dominated by a brighter star ($\Delta V=0.78$) at a
separation of $1\farcs7$.  In the spectra used to analyze this system,
some of the light from the third component was blocked by the slit,
resulting in a flux ratio of 0.4--0.7 inferred from the broadening
function.  Our procedure yields a somewhat lower flux ratio of 0.23, although the program fits part of the third star as if it were the contact binary.
The fit yields very sharp, well-defined minima in variance as a
function of $T_3$ and $\Delta v_{3}$, but our results are nevertheless
poorly defined, since the initial fit to the contact binary was biased
greatly by the presence of the third star (since it dominates the
spectrum).  As a result, the temperature inferred for the contact
binary is too high, and that for the tertiary too low.  Since we use
the observed flux ratio, however, our inferred tertiary mass and mass
ratio should be accurate.

\paragraph{CK Boo} (II) is an A-type W Uma system.  We find a good fit
to the spectrum, and while there are some systematic residuals, a
clear signature of a third star is present, with a very low flux ratio
$\beta=0.009$.  The fitted temperature is consistent
with that inferred from the flux ratio, $T_{3,\beta}=3900\,$K.  The tertiary
is also detected in the adaptive optics observations described in
\citetalias{prir05}.  The separation is only $0\farcs12$ and hence the
magnitude difference $\Delta K\simeq2.8$ is rather uncertain.  We
nevertheless tried to verify consistency: using
Table~\ref{tab:properties} and \citet[ Tables 7.6, 15.7, and
15.8]{cox00}, we infer $M_K=2.6$ for CK Boo and thus
$M_{K,3}\simeq5.4$.  The latter implies that the tertiary would be an
$\sim\!0.48\,M_\odot$ M1 star with $T_3\simeq3800\,$K, $V-K\simeq3.8$
and $M_{V,3}\simeq9.2$.  The temperature and implied flux ratio
of 0.007 agree well with our measurements.  We note that from
arrival-time variations, in \citetalias{prir05} the possibility of a
companion in an $\sim\!5\,$AU orbit was mentioned.  The inferred
minimum mass of $1.5\,M_\odot$, however, is inconsistent with our
results, unless it were a neutron star or black hole.

\paragraph{FI Boo} (IV) is a W-type contact binary system.  We 
clearly detect a faint third component, with $\beta=0.012$ and
$T_{\rm3,fit}=3900\,$K.  The fit to the binary has relatively large
systematic residuals, which dominate the variance; as a result adding
a third star does not change the variance as much as might be expected
to if the fit were better.  Despite these limitations, there is an
obvious minimum in of the variance as a function of $T_3$ at a much
later spectral type than that of the main binary.  Furthermore, the
variance shows a sharp drop at $\Delta v_{3}\simeq0$.  The presence of
a tertiary is also inferred from stochastic residuals in Hipparcos
measurements \citepalias{prir05}.

\paragraph{SV Cam} (VI) is a detached binary, for which the likely
presence of a third body in a 41 or 58\,yr orbit was inferred from
arrival-time variations \citep{lehhw02,borpc04}; the implied
separation is a few $0\farcs1$ and the minimum mass is around
$0.2\,M_\odot$.  The different temperatures of the binary components
make it less suited to our method of analysis, and our fit to the
average spectrum is relatively poor.  We nevertheless clearly detect a
tertiary, and infer a temperature $T_{\rm3,\beta}=3900\,$K that agrees
nicely with the one derived from the fit, $T_{\rm3,fit}=3900\,$K.  We
note, however, that because of the poor fit, there is a marked
decrease in variance for all tertiary spectral types: likely, the
third star is being fit to some of the residuals left from fitting the
main binary.  As a result, the fitted flux ratio of 0.016 may be
somewhat higher than it would be if it were fitting only the third
star.  Since the temperature is close to the lower limit of our range
of templates, the real temperature may well be lower.  An independent
indication for a lower temperature would be that the inferred mass of
$M_3=0.65\,M_\odot$ is somewhat high compared with that inferred from
the arrival-time orbit: it would require an inclination
$i_3\lesssim20^\circ$, which has an {\em a priori} probability of
$\lesssim\!5$\%.  Furthermore, \citet{lehhw02} mention that ``masses
of $\geq\!0.60\,M_\odot$ should be excluded because a third stellar
spectrum would be visible in the observations which is not the case.''
Since SV Cam is not a contact binary, we do not include it in our
statistical analysis.

\paragraph{V776 Cas} (V), an A-type contact binary, is the brighter
member of a visual binary, with an angular separation of $5\farcs38$
and $\Delta V=1.56$ \citepalias{tok97}.  Our fit to the contact binary
is fair, and easily detect the third star.  Our flux ratio is much
lower than the measured one since most of the tertiary's light fell
outside of the slit (indeed, many of the individual spectra of V776
Cas were taken on purpose excluding the third star to make it easier
to calculate the radial velocity for the individual components).  The
fitted temperature $T_{\rm3,fit}=6100\,$K is somewhat higher than that
inferred from the flux ratio.

\paragraph{EE Cet} (VI) is a component of a visual binary in which the
second component, at $5\farcs6$, is another close binary (hence, the
system is a quadruple).  The contact binary is the fainter component,
by $\Delta V=0.36$ \citepalias{tok97}.  The data used in our
observations were taken excluding as much light from the other binary
as possible, but our program nevertheless picks out scattered light
from the companions and identifies it as a very faint third star.  We
do not list it in Table~\ref{tab:properties}, since without knowing
that the system was a multiple, we would not have identified it as
such.  For reference, we note that our fit yields $T_{12}=6200\,$K and
$T_3=6800\,$K.

\paragraph{V445 Cep} (IX) is an A-type contact binary with very shallow eclipses. It is a hot system -- we measure T$_{12} = 7400\,$ K  for the main binary. We also find a third star in the system, with a $T_{3}= 5600\, $K and relative flux $\beta$ = 0.055. This flux is high enough that one would have expected the tertiary to have been detected in previous surveys, so it is a bit puzzling that it has not. Nonetheless, the fact that there is a clear minimum in the variance when a third star of a very different spectral type is added leads us to conclude that there is a third component in the system.

\paragraph{KR Com} (VI) is an A-type contact binary in a visual
binary, with a companion at $0\farcs119$ that is fainter by $\Delta
V=0.59$, or $\beta=0.58$ \citepalias{tok97}.  From the broadening
function, a flux ratio of 0.56 was inferred, while we find a much
lower value of 0.23, likely because our fit is biased by the fact
that the tertiary is so bright.  Because of this, we use the observed
flux ratio to infer the tertiary's parameters.

\paragraph{V401 Cyg} (VI) is a contact binary for which the spectral
signature of the tertiary, despite being only at the 3\% level, was
already seen in the broadening function.  Our fit to the binary is
poor in a somewhat surprising fashion: in some parts, it reproduces
the spectrum very well, while in others, particularly around
5170\,\AA, it fails utterly.  Despite the resulting uncertainties, we
clearly recover the third star with $\beta=0.015$.  The
temperatures inferred from the fit and the flux ratio are both 4700K. 
The presence of a close-in tertiary is also inferred from stochastic 
residuals in Hipparcos measurements \citepalias{prir05}; 
it cannot be the object at a separation of
$18\farcs0$ in the adaptive optics observations presented in
\citetalias{prir05}.

\paragraph{V2082 Cyg} (IX) is likely an A-type contact binary,
although a detached configuration cannot be completely excluded.  We
measure $T_{12}=7000\,$K, and obtain a fairly good fit, although there
are a number of features -- at 5235, 5195, and 5167\AA\ -- that are
stronger than in the template.  Nevertheless, we are able to detect a
faint third companion, with $\beta=0.02$ and
$T_{\rm3,fit}=5100\,$K (the latter consistent with
$T_{\rm3,\beta}=5200\,$K inferred from the flux ratio).  The only other
indication for the presence of a tertiary found in
\citetalias{prir05}, was that the X-ray flux was stronger than
expected for the early-type binary.

\paragraph{V2150 Cyg} (IV) is an A-type contact binary that has a
much fainter, $\Delta V=3.35$ visual companion at $3\farcs68$
\citepalias{prir05}.  The faint companion star was outside the slit in
our spectra, and unlike for brighter, well-separated visual doubles in
our sample (such as EE Cet), we detect no scattered light from the
third companion, nor do we find any closer companion.  Our detection
limit is at the 1\% level, despite the fact that the binary's spectrum
is not fit all that well, with relatively large, broad-scale
deviations as well as some higher-frequency residuals.  This may
partly be because the temperature of about 7200\,K places V2150 Cyg
near the upper end of our temperature range, where we have relatively
few template spectra.

\paragraph{V899 Her B} (IV) is an A-type contact binary for which the
broadening profiles indicate it is the fainter component of a
spectroscopic triple: $\beta=1.5$.  The brighter star is itself
a radial-velocity variable as well.  From our procedure, we infer a
smaller value for the flux ratio, 0.725, but this is likely because the
initial fit to the spectrum is biased by the dominating flux from the
third star.  Hence, the temperature for the binary is overestimated.
Nevertheless, the overall fit is fairly good.  In \citetalias{prir05},
no other evidence for the presence of a tertiary is listed.

\paragraph{ET Leo} (VI) is a low-amplitude contact binary presumably
seen at low inclination.  Our initial fit to the binary spectrum left
rather strong, large-scale residuals; these were reduced but not
altogether removed using a tenth-order polynomial fit to the
continuum.  Our best-fit temperature for the binary is 5800\,K.  This
is substantially higher than what would be inferred from the spectral
type of G8 assigned by \citet{rvddo-vi}, but consistent with the
temperature inferred from $B-V$.  The residuals from the best fit show
a clear signature of an M-type tertiary, and this is confirmed by the
variance as a function of $T_3$ and $\Delta v_3$.  The inferred flux
ratio and temperature are $\beta=0.022$ and
$T_{\rm3,fit}=3900\,$K.  The presence of the tertiary was
also suspected based on residuals of Hipparcos measurements
\citepalias{prir05}.

\paragraph{VZ Lib} (IV) is a contact binary for which the presence of
a fainter tertiary component was already indicated by its signature in
the broadening function.  No other indicators for a tertiary were
found in \citetalias{prir05}.  We recover the third star
unambiguously, but find $\beta=0.045$, which is much fainter
than the value of 0.2 estimated from the broadening function.  This
may be because the previous measurement effectively measured the
tertiary's flux assuming it had the same spectral type as the binary,
while in reality it is later and hence has stronger lines
(Sect.~\ref{sec:previous}).  Our overall fit is good, although some of
the sharp-line features (particularly at 5182\,\AA) are reproduced
relatively poorly, indicating that the tertiary spectral type or
metallicity may not be entirely correct.  The tertiary's temperature estimates also not quite in agreement, T$_{3,\beta}=4200\,$K from the flux ratio and T$_{3,fit}=4700\,$K from the temperature.

\paragraph{SW Lyn} (IV) is a close, semi-detached binary in a
`reversed-Algol' configuration, with the more massive component
filling its Roche lobe.  Despite the fact that the components are not
in contact and therefore have different temperatures, we easily
recover the tertiary that was identified from the broadening function.
We find $\beta=0.19$, substantially less than the value of 0.33
inferred before (likely because the latter does not take into account
that the tertiary has a different spectral type;
Sect.~\ref{sec:previous}).  Overall, the fit is good, although some
mismatches remain.  Since this system is not a contact binary, it was
not used in the statistical analysis.

\paragraph{V502 Oph} (IX) is a W-type contact binary, for which there
are several pieces of evidence pointing to a third companion.  First,
\citet{hugmcl84} detected two radio sources near the source, separated
by only $2\farcs6$.  Second, \citet{derd92} found that the arrival
times of the minima showed a modulation with a period of about 35
years (which, however, is too short for a companion separated by
$2\farcs6$).  Third, \cite{henm98} detected stationary \ion{Na}{1}
lines in trailed spectra.  The latter detection is the most
convincing, and is the basis of the identification as a triple in
\citetalias{prir05}.  Unexpectedly, our procedure does not
unambiguously recover the tertiary.  We find a reasonable fit for a
binary temperature T$_{\rm12,fit}=5800\,$K, but with some rather odd
residuals throughout the spectrum, indicating our template is not a
good match.  Adding a third star, we do see an improvement in the
quality of the fit, but the minimum in variance does not occur close
to zero relative velocity.  Furthermore, taken at face value, the
relative flux of $\beta=0.007$ indicates a tertiary temperature
much lower than the fitted value, $T_{\rm3,fit}=6900\,$K.  For this
reason, we have not counted this system as a detection.  We note,
however, that the discrepancy might be reduced if the separation is
really $2\farcs6$, since in that case much of the tertiary's light
might have fallen outside the slit.

\paragraph{V2388 Oph} (VI) is a W UMa member of a bright visual
binary, with a separation of 0$\farcs$087 and magnitude difference
$\Delta V=1.75$ \citepalias{tok97}.  It is possibly seen in
arrival-time and astrometric variations as well \citepalias{prir05}. 
From the broadening function, a tertiary flux ratio $\beta=0.2$
was found, while our procedure yields a value of~0.10.  We find we can
reproduce the spectrum very well, although the initial fit (without a
third star) largely incorporates the light from the third star.
Hence, the inferred binary temperature, $T_{\rm12,fit}=6100\,$K, will
be biased somewhat (the fact that the temperature inferred from $B-V$
is similar likely reflects the fact that the colour is contaminated by
the tertiary as well).  Nonetheless, we find a very different spectral
type for the tertiary, with $T_{\rm3,fit}=5900\,$K, consistent with
what is inferred from the flux ratio.

\paragraph{BB Peg} (I) is a W-type contact binary in which the
combination of a light-curve fit and radial-velocity orbits allowed
the masses of both components to be measured: $M_1= 1.38\,M_\odot$ and
$M_2=0.5\,M_\odot$.  The primary mass is in good agreement with the
mass inferred from the absolute magnitude through the
period-luminosity-color relation.  The average spectrum, which is
among the more noisy we analyzed, is reproduced fairly well by our
procedure, although some small deviations on relatively large scales
remain.  There is a clear drop in the variance as cooler third stars
are added to the system, indicating the presence of a third component
in the system with a relative flux $\beta=0.009$ and temperature
$T_{\rm3,fit}=3900\,$K.  A third component is also suspected from
arrival-time variations \citepalias{prir05}.

\paragraph{KP Peg} (IX) is a $\beta$-Lyrae type binary, and is the
brighter component of a visual binary with a separation of $3\farcs5$
and $\Delta V=1.6$.  It has one of the earliest spectral types, A2,
implying a temperature that is outside the range covered by our
templates.  Nevertheless, we find a fairly good fit to the binary, and
easily detect the third star.  We find $\beta=0.031$, which is
dimmer than the known flux ratio since most of the light of the third
star did not enter the slit.  The fitted tertiary temperature of
$T_{\rm3,fit}=7700\,$K is not consistent with what is expected for the
observed magnitude difference.

\paragraph{V335 Peg} (IX) is an A-type contact binary with the second component contributing only 5\% of the total flux. We measure T$_{12} = 6400\,$ K, and see some minor residuals. The variance profile for the third star declines for later type stars, indicating a third star with $T_{3} = 3700\,$K. The estimated flux it finds for the third star is only $\beta=$0.006, however, which is right at our detection limit. Hence, we classify it as an interesting non-detection.

\paragraph{HX UMa} (VIII) is a A-type contact binary with a previously
identified, fainter ($\Delta V=3.31$) companion at a separation
of~$0\farcs626$ \citepalias{tok97}.  The tertiary's signal was also
detected in the broadening function, yielding $\beta=0.049$.  Our
procedure provides an excellent fit to the main binary, with the fit
improving even further with the addition of a third star.  We find
$\beta=0.023$, with is somewhat lower than indicated by the magnitude difference and
inconsistent with the measurement from the broadening function.  It is
possible that this results partly from light not entering the slit and
partly from the bandpass being blueward of~$V$.  (If so, the good
agreement found earlier would be due to a fortuitous cancellation of
the light loss by the overestimate of the flux resulting from the
assumption that the tertiary had the same spectral type as the binary;
Sect.~\ref{sec:previous}.)  The fitted tertiary temperature is
$T_{\rm3,fit}=4400\,$K, which is in excellent agreement with that
inferred from the observed flux ratio.

\paragraph{II UMa} (VI) is an A-type contact binary that has a
fainter, $\Delta V=1.64$, companion at a separation of $0\farcs87$
\citepalias{tok97}.  The tertiary was obvious already in the
broadening function, yielding $\beta=0.17$, and our procedure
easily recovers it.  We find $\beta=0.15$.  Unlike other cases,
this is consistent with the value found from the broadening function,
since the spectral types of the contact binary and the tertiary are
very similar.

\paragraph{HT Vir B} (IV) is part of a close visual binary, with a period
of 274\,yr and semi-major axis of $1\farcs01$ \citep{hei86}, current
separation $\sim\!0\farcs6$), and the spectrum shows strong lines from
the third star.  The contact binary is brighter than the companion
during its maxima ($\Delta V=0.63$), but fainter during the minima.
\citet{rvddo-iv} found radial-velocity variations in the third
component, indicating that it is in a binary itself and hence that the system is
a quadruple.  Our procedure yields a flux ratio $\beta=0.28$,
lower than the value of 0.52 inferred from the broadening
function. Our fit is fairly good, and we infer temperatures around 6000\,K for
both components, consistent with the values inferred from the colours
and flux ratio.

\bibliographystyle{apj}
\bibliography{triples}

\end{document}